\documentclass[10pt,conference]{IEEEtran}

\usepackage{amsfonts,graphicx,subfigure}
\usepackage{amsfonts,amsmath,amssymb}
\usepackage{epic,eepic,eepicemu}
\usepackage{epsf}
\usepackage{epsfig}
\usepackage{fancyheadings}
\usepackage{graphics}
\usepackage{psfrag}

\makeatletter
\long\def\@makecaption#1#2{
        \vskip 0.8ex
        \setbox\@tempboxa\hbox{\small {\bf #1:} #2}
        \parindent 1.5em  
        \dimen0=\hsize
        \advance\dimen0 by -3em
        \ifdim \wd\@tempboxa >\dimen0
                \hbox to \hsize{
                        \parindent 0em
                        \hfil 
                        \parbox{\dimen0}{\def\baselinestretch{0.96}\small
                                {\bf #1.} #2
                                }                         \hfil}
        \else \hbox to \hsize{\hfil \box\@tempboxa \hfil}
        \fi
        }
\makeatother

\usepackage{ifthen}

\makeatletter
\long\def\@makecaption#1#2{
        \vskip 0.8ex
        \setbox\@tempboxa\hbox{\small {\bf #1:} #2}
        \parindent 1.5em  
        \dimen0=\hsize
        \advance\dimen0 by -3em
        \ifdim \wd\@tempboxa >\dimen0
                \hbox to \hsize{
                        \parindent 0em
                        \hfil 
                        \parbox{\dimen0}{\def\baselinestretch{0.96}\small
                                {\bf #1.} #2
                                } 
                        \hfil}
        \else \hbox to \hsize{\hfil \box\@tempboxa \hfil}
        \fi
        }
\makeatother

\long\def\comment#1{}

\makeatletter
\def\@cite#1#2{[\if@tempswa #2 \fi #1]}
\makeatother

\makeatletter
\long\def\barenote#1{
    \insert\footins{\footnotesize
    \interlinepenalty\interfootnotelinepenalty 
    \splittopskip\footnotesep
    \splitmaxdepth \dp\strutbox \floatingpenalty \@MM
    \hsize\columnwidth \@parboxrestore
    {\rule{\z@}{\footnotesep}\ignorespaces
      #1\strut}}}
\makeatother

\newcommand{\bit}{\begin{itemize}}
\newcommand{\eit}{\end{itemize}}
\newcommand{\ben}{\begin{enumerate}}
\newcommand{\een}{\end{enumerate}}

\newcommand{\bear}{\begin{eqnarray}}
\newcommand{\eear}{\end{eqnarray}}

\newcommand{\fn}{\footnotesize}

\newcommand{\mess}{\ensuremath{M}}

\newcommand{\compat}{\ensuremath{\psi}}

\newcommand{\graph}{\ensuremath{G}}

\newcommand{\vertex}{\ensuremath{V}}
\newcommand{\edge}{\ensuremath{E}}

\newcommand{\bk}{\ensuremath{\backslash}}

\newcommand{\Exs}{\ensuremath{{\mathbb{E}}}}

\newcommand{\beq}{\begin{quotation}}
\newcommand{\enq}{\end{quotation}}

\newcommand{\estart}{\begin{equation}}
\newcommand{\eend}{\end{equation}}

\newcommand{\widgraph}[2]{\includegraphics[keepaspectratio,width=#1]{#2}}

\newcommand{\defn}{\ensuremath{:  =}}

\newcommand{\bec}{\begin{center}}
\newcommand{\enc}{\end{center}}

\newcommand{\beit}{\begin{itemize}}
\newcommand{\enit}{\end{itemize}}

\newcommand{\been}{\begin{enumerate}}
\newcommand{\enen}{\end{enumerate}}

\newcommand{\comsl}{\begin{slide}}
\newcommand{\comspor}{\begin{slide*}}

\newcommand{\comsld}[2]{\begin{slide}[#1,#2]}
\newcommand{\comspord}[2]{\begin{slide*}[#1,#2]}

\newcommand{\mendsl}{\end{slide}}
\newcommand{\mendspo}{\end{slide*}}

\newcommand{\estim}[1]{\ensuremath{\widehat{#1}}}

\newtheorem{definition}{Definition}
\newtheorem{lemma}{Lemma}
\newtheorem{proposition}{Pemma}

\newcommand{\Ber}{\operatorname{Ber}}

\newcommand{\Checks}{\ensuremath{C}}
\newcommand{\Infobit}{\ensuremath{V}}

\newcommand{\Infoplus}{\ensuremath{\bar{\Infobit}}}

\newcommand{\Genmat}{\ensuremath{A}}

\newcommand{\mypar}{\ensuremath{P}}
\newcommand{\myparo}{\ensuremath{\mypar^1}}
\newcommand{\myparz}{\ensuremath{\mypar^0}}

\newcommand{\Parset}{\ensuremath{\mathcal{P}}}

\newcommand{\lam}{\ensuremath{\lambda}}
\newcommand{\lamo}{\ensuremath{\lam^1}}
\newcommand{\lamz}{\ensuremath{\lam^0}}

\newcommand{\pseumarg}{\ensuremath{\mu}}

\newcommand{\nsou}{\ensuremath{n}} 
\newcommand{\minfo}{\ensuremath{m}}
\newcommand{\Ysou}{\ensuremath{Y}}

\newcommand{\ysou}{\ensuremath{y}}
\newcommand{\zsou}{\ensuremath{z}}
\newcommand{\xinfo}{\ensuremath{x}}

\newcommand{\dham}{\ensuremath{d_H}} 
\newcommand{\rate}{\ensuremath{R}} 
\newcommand{\Code}{\ensuremath{\mathbb{C}}}
\newcommand{\Extcode}{\ensuremath{\bar{\mathbb{C}}}}
\newcommand{\GF}{\ensuremath{\operatorname{GF}}}
\newcommand{\betapar}{\ensuremath{\gamma}}

\newcommand{\Mezard}{M\'{e}zard$\;$}

\newcommand{\messfzero}[2]{\ensuremath{\mess^{0f}_{#1 \rightarrow #2}}}
\newcommand{\messfone}[2]{\ensuremath{\mess^{1f}_{#1 \rightarrow #2}}}
\newcommand{\messwzero}[2]{\ensuremath{\mess^{0w}_{#1 \rightarrow #2}}}
\newcommand{\messwone}[2]{\ensuremath{\mess^{1w}_{#1 \rightarrow #2}}}
\newcommand{\messstar}[2]{\ensuremath{\mess^{*}_{#1 \rightarrow #2}}}

\newcommand{\Bias}{\ensuremath{B}}

\newcommand{\nsoustar}{\ensuremath{n_\ast^{\operatorname{sou}}}}
\newcommand{\ninfostar}{\ensuremath{n_\ast^{\operatorname{info}}}}
\newcommand{\wsou}{\ensuremath{w_{\operatorname{sou}}}}
\newcommand{\winfo}{\ensuremath{w_{\operatorname{info}}}}

\begin{document}

\title{{\huge{Lossy source encoding via message-passing and decimation
over generalized codewords of LDGM codes}}}

\author{\authorblockN{Martin J. Wainwright}
\authorblockA{Departments of EECS and Statistics,
UC Berkeley\\
Berkeley, CA,  94720 \\
Email: wainwrig@eecs.berkeley.edu}
\and
\authorblockN{Elitza Maneva}
\authorblockA{Computer Science Division, 
UC Berkeley\\
Berkeley, CA,  94720 \\
Email: elitza@eecs.berkeley.edu}
}

%

\maketitle

\begin{abstract}
We describe message-passing and decimation approaches for lossy source
coding using low-density generator matrix (LDGM) codes.  In
particular, this paper addresses the problem of encoding a
Bernoulli($\frac{1}{2}$) source: for randomly generated LDGM codes
with suitably irregular degree distributions, our methods yield
performance very close to the rate distortion limit over a range of
rates.  Our approach is inspired by the survey propagation (SP)
algorithm, originally developed by \Mezard et al.~\cite{MPZ02} for
solving random satisfiability problems.  Previous work by Maneva et
al.~\cite{MMWSoda05} shows how SP can be understood as belief
propagation (BP) for an alternative representation of satisfiability
problems.  In analogy to this connection, our approach is to define a
family of Markov random fields over generalized codewords, from which
local message-passing rules can be derived in the standard way.  The
overall source encoding method is based on message-passing, setting a
subset of bits to their preferred values (decimation), and reducing
the code.
\end{abstract}

\begin{center}
\vspace*{-5in}
\vbox to 5in{\small\tt%
 \begin{tabular}[t]{c}
    $\qquad$ To appear in International Symposium on Information Theory;
    Adelaide, Australia
  \end{tabular} \vfil}

\end{center}

\section{Introduction}

Graphical codes such as turbo and low-density parity check (LDPC)
codes, when decoded with the belief propagation or sum-product
algorithm, perform close to capacity~\cite[e.g.,]{Richardson01b}.
Similarly, LDPC codes have been successfully used for various types of
lossless compression schemes~\cite[e.g.,]{Caire03}.  One standard
approach to lossy source coding is based on trellis codes and the
Viterbi algorithm.  The goal of this work is to explore the use of
codes based on graphs with cycles, whose potential has not yet been
fully realized for lossy compression.  A major challenge in applying
such graphical codes to lossy compression is the lack of practical
(i.e., computationally efficient) algorithms for encoding and
decoding.  Accordingly, our focus is the development of practical
algorithms for performing lossy source compression.  For concreteness,
we focus on the problem of quantizing a Bernoulli source with $p =
\frac{1}{2}$.  Developing and analyzing effective algorithms for this
problem is a natural first step towards solving more general lossy
compression problems (e.g., involving continuous sources or memory).

Our approach to lossy source coding is based on the dual codes of LDPC
codes, known as low-density generator matrix (LDGM) codes.  This
choice was partly motivated by an earlier paper of Martinian and
Yedidia~\cite{Martinian03}, who considered a source coding ``dual'' of
the BEC channel coding problem.  They proved that optimal
rate-distortion performance for this problem can be achieved using the
LDGM duals of capacity-achieving LDPC codes, and a modified
message-passing algorithm.  Our work was also inspired by the original
survey propagation algorithm~\cite{MPZ02}, and subsequent analysis by
Maneva et al.~\cite{MMWSoda05} making a precise connection to belief
propagation over an extended Markov random field. In recent work,
Murayama~\cite{Murayama04} developed a modified form of belief
propagation based on the TAP approximation, and provided results in
application to source encoding for LDGM codes with fixed check degree
two.  In work performed in parallel to the work described here, other
research groups have applied forms of survey propagation for source
encoding based on codes composed of local non-linear ``check''
functions~\cite{Mezard05} and $k$-SAT problems with
doping~\cite{Battaglia04}.

\section{Background and set-up}

Given a $\Ber(\frac{1}{2})$ source, any particular i.i.d. realization
$\ysou \in \{0,1 \}^\nsou$ is referred to as a \emph{source sequence}.
The goal is to compress source sequences $\ysou$ by mapping them to
shorter binary vectors $\xinfo \in \{0,1\}^\minfo$ with $\minfo <
\nsou$, where the quantity $\rate \defn \frac{\minfo}{\nsou}$ is the
compression ratio.  The source decoder then maps the compressed
sequence $\xinfo$ to a reconstructed source sequence $\estim{\ysou}$.
For a given pair $(\ysou, \estim{\ysou})$, the reconstruction fidelity
is measured by the Hamming distortion $\dham(\ysou, \estim{\ysou})
\defn \frac{1}{\nsou} \sum_{i=1}^\nsou |\ysou_i - \estim{\ysou}_i|$.
The overall quality of our encoder-decoder pair is measured by the
average Hamming distortion $D \defn \Exs [\dham(\Ysou,
\estim{\Ysou})]$.  For the $\Ber(\frac{1}{2})$ source, the rate
distortion function is well-known to take the form $R(D) = 1 - H(D)$
for $D \in [0, 0.5]$, and $0$ otherwise.

Our approach to lossy source coding is based on low-density generator
matrix codes, hereafter referred to as LDGM codes, which arise
naturally as the duals of LDPC codes.  For a given rate $R =
\frac{m}{n} < 1$, let $\Genmat$ be an $\nsou \times \minfo$ matrix
with $\{0,1\}$ entries, where we assume $\operatorname{rank}\Genmat =
m$ without loss of generality.  The low-density condition requires
that the number of $1$s in each row and column is bounded.  The matrix
$A$ is the generator matrix of the LDGM, thereby defining the code
$\Code(\Genmat) \defn \{ \zsou \in \{0,1\}^\nsou \, | \, \zsou =
\Genmat \xinfo \mbox{ for some $\xinfo \in \{0,1\}^m$} \}$, where
arithmetic is performed over $\GF(2)$.  It will also be useful to
consider the code over $(x,z)$ given by $\Extcode(\Genmat) \defn \{
(\xinfo, \zsou) \in \{0,1\}^{\nsou + \minfo} \: | \; \zsou = \Genmat
\xinfo \}$.  We refer to elements of $\xinfo$ as \emph{information
bits}, and elements of $\zsou$ as \emph{source bits}.  In the LDGM
approach to source coding, the encoding phase of the source coding
problem amounts to mapping a given source sequence $\ysou \in
\{0,1\}^\nsou$ to an information vector $\xinfo(\ysou) \in
\{0,1\}^\minfo$.  Decoding is straightforward: we simply form
$\estim{\ysou}(\xinfo) = \Genmat \xinfo$.  The challenge lies in the
encoding phase: in particular, we must determine the information bit
vector $\xinfo$ such that the Hamming distortion $\frac{1}{\nsou} \|
\ysou - \Genmat \xinfo \|_1$ is minimized.  This combinatorial
optimization problem is equivalent to an MAX-XORSAT problem, and hence
known to be NP-hard in general.

It is convenient to represent a given LDGM code, specified by generator
matrix $\Genmat$, as a factor graph $\graph = (\Infobit, \Checks,
\edge)$, where $\Infobit = \{1,\ldots, \minfo \}$ denotes the set of
information bits and $\Checks \defn \{1,\ldots, \nsou \}$ denotes the
set of checks (or equivalently, source bits), and $\edge$ denotes the
set of edges between checks and information bits. As illustrated in
Figure~\ref{FigCases}, the $\nsou$ source bits are lined up at the top
of the graph, and each is connected to a unique check neighbor.  Each
check, in turn, is connected to (some subset of) the $\minfo$
information bits at the bottom of the graph.  Note that there is a
one-to-one correspondence between source bits and checks.  We use
letters $a,b,c$ to refer to elements of $\Checks$, corresponding either
to a source bit or the associated check.  Conversely, we use letters
$i,j,k$ to refer to information bits in the set $\Infobit$.  For each
information bit $i \in \Infobit$, let $\Checks(i) \subseteq \Checks$
denote its check neighbors: $\Checks(i) \defn \{ a \in \Checks \: | \;
(a,i) \in \edge \}$.  Similarly, for each check $a \in \Checks$, we
define the set $\Infobit(a) \defn \{ i \in \Infobit \: | \; (a,i) \in
\edge \}$.  We use the notation $\Infoplus(a) \defn \Infobit(a) \cup
\{a\}$ to denote the set of \emph{all} bits---both information and
source---that are adjacent to check $a$.

\section{Markov random fields and decimation with generalized codewords}

A natural first idea to solving the source encoding problem would be
to follow the channel coding approach: run the sum-product algorithm
on the ordinary factor graph, and then threshold the resulting
log-likelihood ratios (LLRs) at each bit to determine a source
encoding $\xinfo(\estim{\ysou})$.  Unfortunately, this approach fails:
either the algorithm fails to converge or the LLRs fail to yield
reliable information, resulting in a poor source encoding.  Inspired
by survey propagation for satisfiability problems~\cite{MPZ02}, we
consider an approach with two components: (a) extending the factor
distribution so as to include not just ordinary codewords but also a
set of partially assigned codewords, and (b) performing a sequence of
message-passing and decimation steps, each of which entails setting
fraction of bits to their preferred values.

More specifically, we consider Markov random fields over a larger
space of so-called generalized codewords, which are members of the
space $\{0,1, \ast\}^{\nsou + \minfo}$ where $\ast$ is a new symbol.
As we will see, the interpretation of $\xinfo_i = \ast$ is that the
associated bit $i$ is \emph{free}.  Conversely, any bit for which
$\xinfo_i \in \{0,1 \}$ is \emph{forced}.  One possible view of a
generalized codeword, as with the survey propagation and $k$-SAT
problems, is as an index for a cluster of ordinary codewords.  We
define a family of Markov random fields, parameterized by a weight for
$\ast$-variables, and a weight that measures fidelity to the source
sequence.  As a particular case, our family of MRFs includes a
weighted distribution over the set of ordinary codewords.  Although
the specific extension considered here is natural to us (and yields
good source coding results), it could be worthwhile to consider
alternative ways in which to extend the original distribution to
generalized codewords.

\subsection{Generalized codewords}
\begin{figure}[h]
\bec
\psfrag{#i#}{$i$}
\psfrag{#j#}{$j$}
\psfrag{#k#}{$k$}
\psfrag{#l#}{$l$}
\psfrag{#a#}{$a$}
\psfrag{#b#}{$b$}
\psfrag{#c#}{$c$}
\psfrag{#d#}{$d$}
\psfrag{#e#}{$e$}
\psfrag{#f#}{$f$}
\psfrag{#g#}{$g$}
\psfrag{#*#}{$\ast$}
\psfrag{#0#}{$0$}
\psfrag{#1#}{$1$}
\widgraph{0.32\textwidth}{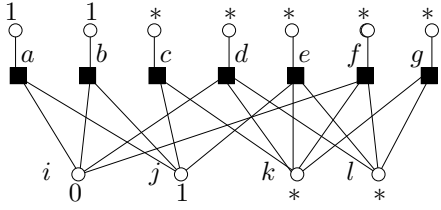}
\enc
\caption{Illustration of a generalized codeword for a small LDGM.
Information bits $i$ and $j$ are both forced; for each, the two
forcing checks are $a$ and $b$.  The remaining checks and bits are all
free.  }
\label{FigCases}
\end{figure}

\begin{definition}[Check states]
\label{DefCheckStates}
In any generalized codeword, each check is in one of two possible
exclusive states: \ben
\item[(i)] we say that check $a \in \Checks$ is forcing whenever none
of its bit neighbors are free, and the local $\{0,1\}$-codeword
$(\zsou_a; \xinfo_{\Infobit(a)}) \in \{0,1\}^{1 + |\Infobit(a)|}$
satisfies parity check $a$.
\item[(ii)] on the other hand, check $a$ is free whenever $\zsou_a =
\ast$, and moreover $\xinfo_i = \ast$ for at least one $i \in
\Infobit(a)$.
\een
\end{definition}
Note that the source bit $\zsou_a$ is free (or forced) if and only if
the associated check $a$ is free (or forcing).  With this set-up, our
space of generalized codewords is defined as follows:
\begin{definition}[Generalized codeword]
\label{DefGenCode}
A vector $(\zsou, \xinfo) \in \{0,1,\ast\}^{\nsou + \minfo}$ is a
valid generalized codeword when the following conditions hold:
\ben
\item[(i)] all checks $a$ are either forcing or free.
\item[(ii)] if some information bit $x_i$ is forced (i.e., $x_i \in
\{0,1\}$), then at \emph{at least} two check neighbors $a \in
\Checks(i)$ must be forcing it.
\een
\end{definition}
For a generator matrix in which every information bit has degree two
or greater, it can be seen that any ordinary codeword $(\zsou, \xinfo)
\in \Extcode(\Genmat)$ is also a generalized codeword.  In addition,
there are generalized codewords that include $\ast$'s in some
positions, and hence do not correspond to ordinary codewords.  One
such (non-trivial) generalized codeword is illustrated in
Figure~\ref{FigCases}.  A natural way in which to generate generalized
codewords is via an iterative ``peeling'' or ``leaf-stripping''
procedure.  Related procedures have been analyzed in the context of
satisfiability problems~\cite{MMWSoda05}, XORSAT
problems~\cite{MezRicZec02}, and for performing binary erasure
quantization~\cite{Martinian03}.

\noindent {\bf{Peeling procedure:}} Given some initial source sequence
$\zsou \in \{0,1, \ast \}^\nsou$, initialize all information bits
$x_i$ to be forced.
\ben
\item While there exists a forced information bit $x_i$ with exactly
one forcing check neighbor $a$, set $x_i = z_a = \ast$.
\item When all remaining forced information bits have at least two
forcing checks, go to Step 3.
\item For any free check $\zsou_a = \ast$ with \emph{no} free
information bit neighbors, set $\zsou_a = \oplus_{i \in \Infobit(a)}
x_i$.
\een
When initialized with at least one free check, Step 1 of this peeling
procedure can terminate in one of two possible ways: either the
initial configuration is stripped down to the all-$\ast$
configuration, or Step 1 terminates at a configuration such that every
forced information bit has two or more forcing check neighbors, thus
ensuring that condition (ii) of Definition~\ref{DefGenCode} is
satisfied.  As noted previously~\cite{Martinian03}, these cores can be
viewed as ``duals'' to stopping sets in the dual LDPC.  Finally, Step
3 ensures that every free check has at least one free information bit,
thereby satisfying condition (i) of Definition~\ref{DefGenCode}.

\subsection{Weighted version}
Given a particular source sequence $y \in \{0,1\}^\nsou$, we form a
probability distribution over the set of generalized codewords as
follows.  For any generalized codeword $(\zsou, \xinfo) \in
\{0,1,\ast\}^{\nsou + \minfo}$, we define the sets
\mbox{$\nsoustar(\zsou) \defn \big | \{ i \in \{1, \ldots, \nsou\} \,
| \, z_i = \ast \} \big |$} and \mbox{$\ninfostar(\xinfo) \defn \big |
\{ i \in \{1, \ldots, \minfo\} \, | \, x_i = \ast \} \big |$,}
corresponding to the number of $\ast$-variables in the source and
information bits respectively.  We associate non-negative weights
$\wsou$ and $\winfo$ with the $\ast$-variables in the source and
information bits respectively.  Finally, we introduce a non-negative
parameter $\gamma$, which will be used to penalize disagreements
between the source bits $\zsou$ and the given (fixed) source sequence
$\ysou$.  Of interest to us in the sequel is the weighted probability
distribution
\begin{equation}
\label{EqnWeighted}
p(\zsou, \xinfo; \wsou, \winfo, \lam) \propto \wsou^{\nsoustar(\zsou)}
\times \winfo^{\ninfostar(\xinfo)} \times \exp^{-2 \gamma \dham(\ysou,
\zsou)}.
\end{equation}
Note that for $\wsou = \winfo = 0$, this distribution reduces to the
standard weighted distribution over ordinary codewords.

\subsection{Representation as Markov random field}

We now seek to represent the set of generalized codewords as a Markov
random field (MRF).  A first important observation is that state
augmentation is necessary to achieve such a Markov representation with
respect to the original factor graph.

\begin{lemma}
For positive $\wsou, \winfo$, the set of generalized codewords
\emph{cannot} be represented as a Markov random field based on the
original factor graph $\graph$ where the state space at each bit is
simply $\{0,1, \ast \}$.
\end{lemma}
\begin{proof}
It suffices to demonstrate that it is impossible to construct an
indicator function for membership in the set of generalized codewords
as a product of local compatibility functions on $\{0,1, \ast \}$, one
for each check.  The key is that the set of all local generalized
codewords cannot be defined only in terms of the variables
$x_{\Infoplus(a)}$; rather, the validity depends also on all bit
neighbors of checks that are incident to bits in $\Infoplus(a)$.  or
more formally on bits with indices in the set
\begin{equation}
\label{EqnExtSet}
\cup_{i \in \Infoplus(a) } \big \{ j \in \Infobit \; \big | \; j \in
\Infobit(b) \mbox{ for some } b \in \Checks(i) \big \}.
\end{equation}

As a particular illustration, consider the trivial LDGM code
consisting of a single source bit (and check) connected to three
information bits.  From Definition~\ref{DefCheckStates} and
Definition~\ref{DefGenCode}, it can be seen that the only generalized
codeword is the all-$\ast$ configuration.  Thus, any check function
used to define membership in the set of generalized codewords would
have to assign zero mass to any other $\{0,1,\ast\}$ configuration.
Now suppose that this simple LDGM is embedded within a larger LDGM
code.  For instance, consider the check labeled $e$ (with source bit
$\zsou_e$) and corresponding information bits $\{j,k, l\}$ in
Figure~\ref{FigCases}.  With respect to the generalized codeword in
this figure, we see that the local configuration $(x_j, x_k, x_l
\zsou_e) = (1, \ast, \ast, \ast)$ is locally valid, which contradicts
our conclusion from considering the trivial LDGM code in isolation.
Hence, the constraints enforced by a given check change depending on
the larger context in which it is embedded.
\end{proof}
\hfill 
%

Consequently, obtaining a factorization of the distribution requires
keeping track of variables in the extended set~\eqref{EqnExtSet}.
Accordingly, as in the reformulation of survey propagation for SAT
problems by Maneva et al.~\cite{MMWSoda05}, we introduce a new
variable $\mypar_i$, so that there is a vector $(x_i, \mypar_i)$
associated with each bit.  To define $\mypar_i$, first let $\Parset(i)
= \Parset(\Checks(i))$ denote the power set of all of the clause
neighbors $\Checks(i)$ of bit $i$.  (I.e., $\Parset(i)$ is a set with
$2^{|\Checks(i)|}$ elements).  The variable $\mypar_i$ takes on
subsets of $\Checks(i)$, and we decompose it as $\mypar_i = \myparz_i
\cup \myparo_i$, where at any time \emph{at most one} of $\myparo_i$
and $\myparz_i$ are non-empty.  The variable $\mypar_i$ has the
following decomposition and interpretation: (a) if $\myparz_i =
\myparo_i = \emptyset$, then no checks are forcing bit $x_i$; (b) if
$\mypar_i = \myparo_i \neq \emptyset$, then certain checks are forcing
$x_i$ to be one (so that necessarily $x_i = 1$); and (c) similarly, if
$\mypar_i = \myparz_i \neq \emptyset$, then certain checks are forcing
$x_i$ to be zero (so that necessarily $x_i = 0$).  By construction,
this definition excludes the case that both $\myparz_i$ and
$\myparo_i$ non-empty at the same time, so that the state space of
$\mypar_i$ has cardinality $2^{|\Checks(i)|} + 2^{|\Checks(i)|} - 1 \;
= \; 2^{|\Checks(i)| + 1} - 1$.

\begin{figure*}

\framebox[1\textwidth]{
\parbox{.98\textwidth}{ 
%
\noindent {\underline{Bits to checks}} {{\fn
\begin{subequations}
\begin{eqnarray*}
\messfzero{i}{a} & \leftarrow & \lamz_i \Big \{\prod_{b \in \Checks(i)
 \bk \{a\}} \big[ \messfzero{b}{i} + \messwzero{b}{i} \big ] -
 \prod_{b \in \Checks(i) \bk \{a \}} \messwzero{b}{i} \Big \} \\
\messfone{i}{a} & \leftarrow & \lamo_i \Big \{\prod_{b \in \Checks(i)
\bk \{a\}} \big[ \messfone{b}{i} + \messwone{b}{i} \big ] - \prod_{b
\in \Checks(i) \bk \{a\}} \messwone{b}{i} \Big \} \\
\messwzero{i}{a} & \leftarrow & \lamz_i \Big \{\prod_{b \in \Checks(i)
\bk \{a\}} \big[ \messfzero{b}{i} + \messwzero{b}{i} \big ] - \prod_{b
\in \Checks(i) \bk \{a \} } \messwzero{b}{i} - \sum_{c \in \Checks(i)
\bk \{a\}} \messfzero{c}{i} \prod_{b \in \Checks(i) \bk \{a, c\} }
\messwzero{b}{i} \Big \} . \\
\messwone{i}{a} & \leftarrow & \lamo_i \Big \{ \prod_{b \in \Checks(i)
\bk \{a\}} \big[ \messfone{b}{i} + \messwone{b}{i} \big ] - \prod_{b
\in \Checks(i) \bk \{a \} } \messwone{b}{i} - \sum_{c \in \Checks(i)
\bk \{a\}} \messfone{c}{i} \prod_{b \in \Checks(i) \bk \{a, c\} }
\messwone{b}{i} \Big \} . \\
\messstar{i}{a} & \leftarrow & \winfo \; \prod_{b \in \Checks(i) \bk \{a\}}
\messstar{b}{i}
\end{eqnarray*}
\end{subequations}
}}
%
%
\noindent {\underline{Checks to bits}} {{\fn\begin{subequations}
\begin{eqnarray*}
\messfzero{a}{i} & \leftarrow & \frac{1}{2} \Big[\prod_ {j \in
\Infoplus(a) \bk \{i\}} \big(\messfzero{j}{a} + \messfone{j}{a}\big) +
\prod_{j \in \Infoplus(a) \bk \{i\}} \big(\messfzero{j}{a} -
\messfone{j}{a}\big) \Big] \\
\messfone{a}{i} & \leftarrow & \frac{1}{2} \Big[\prod_{j \in
\Infoplus(a) \bk \{i\}} \big(\messfzero{j}{a} + \messfone{j}{a}\big) -
\prod_{j \in \Infoplus(a) \bk \{i\}} \big(\messfzero{j}{a} -
\messfone{j}{a}\big) \Big] \\
\messwzero{a}{i} & \leftarrow & \prod_{j \in \Infoplus(a) \bk \{i\}}
\big[ \messstar{j}{a} + \messwone{j}{a} + \messwzero{j}{a} \big ] -
\prod_{j \in \Infoplus(a) \bk \{i\}} \big[\messwone{j}{a} +
\messwzero{j}{a} \big] - \sum_{k \in \Infoplus(a) \bk \{i\}}
\messstar{k}{a} \prod_{j \in \Infoplus(a) \bk \{i, k\}}
\big[\messwone{j}{a} + \messwzero{j}{a}\big] \\
\messwone{a}{i} & = & \messwzero{a}{i}. \\
\messstar{a}{i} & \leftarrow & \prod_{j \in \Infoplus(a) \bk \{i\}}
\big[ \messstar{j}{a} + \messwone{j}{a} + \messwzero{j}{a} \big] -
\prod_{j \in \Infoplus(a) \bk \{i\}} \big[\messwone{j}{a} +
\messwzero{j}{a} \big]
\end{eqnarray*}
\end{subequations}
}}
%
}}
\caption{ Message-passing updates involve five types of messages from
 bit to check, and five types of messages from check to bit.  Any
 source bit $z_a$ always sends to its only check $a$ the message
 5-vector $(\compat_a(0), \; \compat_a(1), \; 0, \; 0, \;\wsou)$.  The
 message vector in any given direction on any edge is normalized to
 sum to one.}
\label{FigMessPass}
\end{figure*}

\subsection{Compatibility functions}

We now specify a set of compatibility functions to capture the Markov
random field over generalized codewords.

\subsubsection{Variable compatibilities} For each bit index $i$ (or
$a$), let $\lamo_i$ and $\lamz_i$ denote the weights assigned to the
events $\xinfo_i = 1$ and $\xinfo_i = 0$ respectively.  For source
encoding, these weights are specified as $\lamo_i = \lamz_i = 1$ for
all information bits $i$ (i.e., no a priori bias on the information
bits), so that the compatibility function takes the form:
\begin{equation}
\compat_i(x_i, \mypar_i) \defn \begin{cases}
1 & \mbox{ if $x_i = 1$ and $|\mypar_i| = |\myparo_i| \geq 2$ }
\\
1 & \mbox{ if $x_i = 0$ and $|\mypar_i| = |\myparz_i| \geq 2$ } \\
\winfo & \mbox{ if $x_i = \ast$ and $\mypar_i = \emptyset$ } 
				   \end{cases}
\end{equation}
The source bits have compatibility functions of the form
$\compat_a(\zsou_a, \mypar_a) =\lamz_a$ if $\zsou_a = 0$ and
$\myparo_a = \{a\}$; $\compat_a(\zsou_a, \mypar_a) =\lamo_a$ if
$\zsou_a = 1$ and $\myparo_a = \{a\}$; and $\compat_a(\zsou_a,
\mypar_a) = \wsou$ if $\zsou_a = \ast$ and $\mypar_a = \emptyset$.
Here $\lamo_a \defn \ysou_a \exp(\betapar) + (1 - \ysou_a)
\exp(-\betapar)$, $\lamz_a \defn 1/\lamo_a$, and the parameter
$\betapar > 0$ reflects how strongly the source observations are
weighted.

\subsubsection{Check compatibilities}

\newcommand{\checkcom}{\ensuremath{\phi}}

For a given check $a$, the associated compatibility function
$\checkcom_a(x_{\Infobit(a)}, \zsou_a, \mypar_{\Infobit(a)})$ is
constructed to ensure that the following two properties hold: (1) The
configuration $\{\zsou_a\} \cup \xinfo_{\Infobit(a)}$ is \emph{valid}
for check $a$, meaning that (a) either it includes no $\ast$'s, in
which case the pure $\{0,1\}$ configuration must be a local codeword;
or (b) the associated source bit is free (i.e., $\zsou_a = \ast$), and
$\xinfo_i = \ast$ for at least one $i \in \Infobit(a)$.  (2) For each
index $i \in \Infobit(a)$, the following condition holds: (a) either
$a \in \mypar_i$ and $a$ forces $x_i$, or (b) there holds $a \notin
\mypar_i$ and $a$ does not force $x_i$.

\newcommand{\pwei}{\ensuremath{p^{wei}}}

\begin{proposition}
With the singleton and factor compatibilities as above, consider the
distribution $\pwei( (x, \mypar(x)), (z, \mypar(z)))$, defined as a
Markov random field (MRF) over the factor graph in the following way:
\begin{equation}
\label{EqnExtendedMRF}
\prod_{i \in \vertex} \compat_i(x_i, \mypar_i) \; \prod_{a \in
\Checks} \compat_a(z_a, \mypar_a) \checkcom_a(x_{\Infobit(a)},
\zsou_a, \mypar_{\Infobit(a)}).
\end{equation}
Its marginal distribution over $(x, z)$ agrees with the weighted
distribution~\eqref{EqnWeighted}.
\end{proposition}

\subsection{Message-passing updates}

In our extended Markov random field, the random variable at each bit
node $i$ is of the form $(x_i, \mypar_i)$, and belongs to the
Cartesian product $\{0,1, \ast\} \times [\Parset(i) \times \{0,1\}]$.
(To clarify the additional $\{0,1\}$, the variable $\mypar_i =
\myparz_i \cup \myparo_i$ corresponds to a particular subset of
$\Parset(i)$, but we also need to specify whether $\mypar_i =
\myparz_i$ or $\mypar_i = \myparo_i$.)  Although the cardinality of
$\Parset(i)$ can is exponential in the bit degree, it turns out that
message-passing can be implemented by keeping track of only five
numbers for each message (in either direction). These five cases are
the following:
\begin{enumerate}
\item[(i)] {\bf{$(x_i = 0, a \in \myparz_i)$:}} check $a$ is forcing
$x_i$ to be equal to zero.  We say $x_i$ is a \emph{forced zero with
respect to $a$,} and use $\messfzero{i}{a}$ and $\messfzero{a}{i}$ for
the corresponding bit-to-check and check-to-bit messages.
\item[(ii)] {\bf{$(x_i = 1, a \in \myparo_i)$:}} check $a$ is forcing
$x_i$ to be equal to one. We say that $x_i$ is a \emph{forced one with
respect to $a$,} and denote the corresponding messages
$\messfone{i}{a}$ and $\messfone{a}{i}$.

\item[(iii)] {\bf{$(x_i = 0, \emptyset \neq \myparz_i \subseteq
\Checks(i)\bk \{a\})$:}} A check subset \emph{not} including $a$ is
forcing $x_i = 0$. We say $x_i$ is a \emph{weak zero with respect to
check $a$}, and denote the messages $\messwzero{i}{a}$ and
$\messwzero{a}{i}$.
\item[(iv)] {\bf{$(x_i = 1, \emptyset \neq \myparo_i \subseteq
\Checks(i)\bk \{a\})$:}} A check subset \emph{not} including $a$
forces $x_i = 1$.  We say that that $x_i$ is a \emph{weak one with
respect to check $a$,} and use corresponding messages
$\messwone{i}{a}$ and $\messwone{a}{i}$.
\item[(v)] {\bf{$(x_i = \ast, \myparo_i = \myparz_i= \emptyset$):}} No
checks force bit $x_i$; associated messages are denoted by
$\messstar{i}{a}$ and $\messstar{i}{a}$.
\end{enumerate}
The differences between these cases is illustrated in
Figure~\ref{FigCases}.  The information bit $x_i = 0$ is a forced zero
with respect to checks $a$ and $b$ (case (i)), and a weak zero with
respect to checks $d$ and $f$ (case (iii)).  Similarly, the setting
$x_j = 1$ is a forced one for checks $a$ and $b$, and a weak one for
checks $c$ and $e$.  Finally, there are a number of $\ast$ variables
to illustrate case (v).  With these definitions, it is straightforward
(but requiring some calculation) to derive the BP message-passing
updates as applied to the generalized MRF, as shown in
Figure~\ref{FigMessPass}.  It can be seen that this family of
algorithms includes ordinary BP as a special case: in particular, if
$\wsou = \winfo = 0$, then the updates reduce to the usual BP updates
on a weighted MRF over ordinary codewords.

\subsection{Decimation based on pseudomarginals}
When the message updates have converged, the sum-product
pseudomarginals (i.e., approximations to the true marginal
distributions) are calculated as follows:
\begin{eqnarray*}
\pseumarg_i(0) & \propto & \; \lamz_i \Big\{ \prod_{a \in \Checks(i)}
\big[ \messfzero{a}{i} + \messwzero{a}{i} \big] - \prod_{a \in
\Checks(i)} \messwzero{a}{i} \\
& & \qquad \quad - \sum_{b \in \Checks(i)} \messfzero{b}{i} \prod_{a
\in \Checks(i) \bk \{b\}} \messwzero{a}{i} \Big \} \\
%
%
\pseumarg_i(\ast) & \propto & \winfo \prod_{a \in \Checks(i)}
\messstar{a}{i}.
\end{eqnarray*}
with a similar expression for $\pseumarg_i(1)$. The overall triplet is
normalized to sum to one.  As with survey propagation and SAT
problems~\cite{MPZ02,MMWSoda05}, the practical use of these
message-passing updates for source encoding entail: (1) Running the
message-passing algorithm until convergence; (2) Setting a fraction of
information bits, and simplifying the resulting code; and (3) Running
the message-passing algorithm on the simplified code, and repeating.
We choose information bits to set based on bias magnitude
\mbox{$\Bias_i \defn |\pseumarg_i(1) - \pseumarg_i(0)|$.}

\section{Results}
\begin{figure}[h]
\bec \widgraph{.43\textwidth}{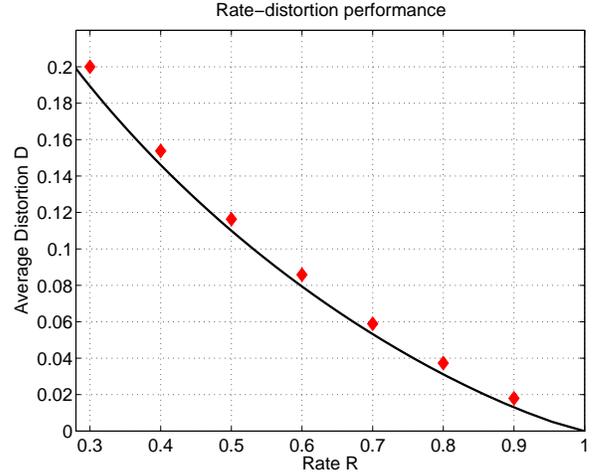}
\caption{Plot of rate versus distortion, comparing the Shannon limit
(solid line) and empirical performance using LDGM codes with
blocklength $\nsou = 10,000$.  Each diamond is the average distortion
over 15 trials.}
\label{FigEmpPerf}
\enc
\end{figure}

We have applied a C-based implementation of our algorithm to LDGM
codes with various degree distributions and source sequences of length
$\nsou$ ranging from $200$ to $100,000$.  Although message-passing can
be slow to build up appreciable biases for regular degree
distributions, we find that biases accumulate quite rapidly for
suitably irregular degree distributions.  We chose codes randomly from
irregular distributions optimized\footnote{This choice, though not
optimized for source encoding, is reasonable in light of the
connection between LDPC channel coding and LDGM source coding in the
erasure case~\cite{Martinian03}.} for ordinary message-passing on the
BEC or BSC using density-evolution~\cite{Richardson01b}.
Figure~\ref{FigEmpPerf} compares experimental results to the
rate-distortion bound $R(D)$. We applied message-passing using a
damping parameter $\alpha = 0.50$, and with $\wsou = 1.10$, $\winfo =
1.0$ and $\betapar$ varying from $1.45$ (for rate $0.90$) to $0.70$
(for rate $0.30$).  Each round of decimation entailed setting all
information bits with biases above a given threshold, up to a maximum
of $2\%$ of the total number of bits.  As seen in
Figure~\ref{FigEmpPerf}, the performance is already very good even for
intermediate block length $\nsou = 10,000$, and it improves for larger
block lengths.  After having refined our decimation procedure, we have
also managed to obtain good source encodings (though currently not
quite as good as Figure~\ref{FigEmpPerf}) using ordinary BP
message-passing (i.e., $\wsou = \winfo = 0$) and decimation; however,
in experiments to date, in which we do not adjust parameters
adaptively during decimation, we have found it difficult to obtain
consistent convergence of ordinary BP (and more generally,
message-passing with $\wsou, \winfo \approx 0$) over all decimation
rounds.  It remains to perform a systematic comparison of the
performance of message-passing/decimation procedures over a range of
parameters $(\wsou, \winfo, \betapar)$ for a meaningful quantitative
comparison.

There remain various open questions suggested by this work. For
instance, an important direction is developing methods for optimizing
LDGM codes, and the choice of parameters in our extended MRFs for
source encoding.  An important practical issue is to investigate the
tradeoff between the conservativeness of the decimation procedure
(i.e., computation time) versus quality of source encoding.  Finally,
the limiting rate-distortion performance of LDGM codes is a
theoretical question that (to the best of our knowledge) remains open.

\section*{Acknowledgment}
Work partially supported by Intel Corporation Equipment Grant
22978. The authors thank Emin Martinian, Marc M\'{e}zard, Andrea
Montanari and Elchanan Mossel for helpful discussions.

\bibliographystyle{IEEEtran.bst}

\begin{thebibliography}{1}
\providecommand{\url}[1]{#1}
\csname url@rmstyle\endcsname
\providecommand{\newblock}{\relax}
\providecommand{\bibinfo}[2]{#2}
\providecommand\BIBentrySTDinterwordspacing{\spaceskip=0pt\relax}
\providecommand\BIBentryALTinterwordstretchfactor{4}
\providecommand\BIBentryALTinterwordspacing{\spaceskip=\fontdimen2\font plus
\BIBentryALTinterwordstretchfactor\fontdimen3\font minus
  \fontdimen4\font\relax}
\providecommand\BIBforeignlanguage[2]{{%
\expandafter\ifx\csname l@#1\endcsname\relax
\typeout{** WARNING: IEEEtran.bst: No hyphenation pattern has been}%
\typeout{** loaded for the language `#1'. Using the pattern for}%
\typeout{** the default language instead.}%
\else
\language=\csname l@#1\endcsname
\fi
#2}}

\bibitem{MPZ02}
M.~M\'{e}zard, G.~Parisi, and R.~Zecchina, ``Analytic and algorithmic solution
  of random satisfiability problems,'' \emph{Science}, vol. 297, 812, 2002.

\bibitem{MMWSoda05}
E.~Maneva, E.~Mossel, and M.~J. Wainwright, ``A new look at survey propagation
  and its generalizations,'' in \emph{Proceedings of the 16th Annual Symposium
  on Discrete Algorithms (SODA)}, 2005, pp. 1089--1098.

\bibitem{Richardson01b}
T.~Richardson, A.~Shokrollahi, and R.~Urbanke, ``Design of capacity-approaching
  irregular low-density parity check codes,'' \emph{IEEE Trans. Info. Theory},
  vol.~47, pp. 619--637, February 2001.

\bibitem{Caire03}
G.~Caire, S.~Shamai, and S.~Verdu, ``A new data compression algorithm for
  sources with memory based on error-correcting codes,'' in \emph{Information
  Theory Workshop}, Paris, France, 2003, pp. 291--295.

\bibitem{Martinian03}
E.~Martinian and J.~Yedidia, ``Iterative quantization using codes on graphs,''
  in \emph{Allerton Conference on Control, Computing, and Communication},
  October 2003.

\bibitem{Murayama04}
T.~Murayama, ``{T}houless-{A}nderson-{P}almer approach for lossy compression,''
  \emph{Physical Review E}, vol.~69, pp. 035\,105(1)--035\,105(4), 2004.

\bibitem{Mezard05}
M.~M\'{e}zard, January 2005, personal communication, Sante Fe Coding Workshop.

\bibitem{Battaglia04}
D.~Battaglia, A.~Braunstein, J.~Chavas, and R.~Zecchina, ``Source coding by
  efficient selection of ground states,'' Tech. Rep., 2004,
  arXiv:cond-mat/0412652 v1.

\bibitem{MezRicZec02}
M.~M\'{e}zard, F.~Ricci-Tersenghi, and R.~Zecchina, ``Alternative solutions to
  diluted p-spin models and {XORSAT} problems,'' \emph{Jour. of Statistical
  Physics}, vol. 111, p. 105, 2002.

\end{thebibliography}

\end{document}